\documentclass[a4paper,11pt]{article}
\usepackage{graphicx}
\usepackage{citesort}
\usepackage{amsfonts}
\usepackage{amssymb}
\usepackage{latexsym}
\usepackage{url}
\usepackage[normalem]{ulem}
\usepackage{color}
\usepackage{ntheorem}

\usepackage{pstricks,pst-node,pst-coil,pst-plot}

\newcommand{\mnotex}[1]
{\protect{\stepcounter{mnotecount}}$^{\mbox{\footnotesize
$
\bullet$\themnotecount}}$ \marginpar{
\raggedright\tiny\em
$\!\!\!\!\!\!\,\bullet$\themnotecount: #1} }

\newcommand{\jamesx}[1]{}
\renewcommand{\jamesx}[1]{{\mnote{{\color{blue}{\bf jg:}
#1} }}}

\newcommand{\cref}[1]{\mbox{{\color{red}FIXME; what is the 4 for?}}4\emph{\ref{#1})}}

\newcommand{\h}[2]{#1\dotfill\ #2\\\ptc{fixme}}

\newcommand{\malph}{s}

 \newcommand{\noplaing}{\tg }
 \newcommand{\noplainmu}{{\widetilde \mu}}
 \newcommand{\noplainK}{\tK }
 \newcommand{\noplainL}{\tL }
 
 \newcommand{\noplainJ}{{\widetilde J}}
 \newcommand{\noplainH}{{\widetilde H} }
 \newcommand{\noplainm}{{\widetilde m} }
 \newcommand{\noplaintheta}{{\widetilde \theta} }
 \newcommand{\noplainD}{\tD }
 \newcommand{\plaing}{g }
 \newcommand{\plainK}{K }
 \newcommand{\plainL}{L }
 \newcommand{\plainR}{R }
 \newcommand{\plaintau}{\tau  }
 \newcommand{\plaindiv}{{\mathrm{ div}} }
 \newcommand{\plaindivvv}{{\plaindiv} }
 \newcommand{\mynottsigmasquare}{\sigma^2}

\newtheorem{theorem}{Theorem}[section]

\newtheorem{Theorem}[theorem]{\sc  Theorem\rm}

\newtheorem{Lemma}[theorem]{\sc Lemma\rm}

\newtheorem{proposition}[theorem]{\sc Proposition\rm}
\newtheorem{Proposition}[theorem]{\sc Proposition\rm}

\newcommand{\divvv}{\mbox{\rm div}}

\newcommand{\hs}{\cH_{\mbox{\scriptsize sing}}}

\newcommand{\beadl}[1]{\begin{deqarr}\label{#1}}
\newcommand{\eeadl}[1]{\arrlabel{#1}\end{deqarr}}%

\def\nz{\ifmmode {I\hskip -3pt N} \else {\hbox {$I\hskip -3pt N$}}\fi}
\def\zz{\ifmmode {Z\hskip -4.8pt Z} \else
       {\hbox {$Z\hskip -4.8pt Z$}}\fi}
\def\qz{\ifmmode {Q\hskip -5.0pt\vrule height6.0pt depth 0pt
       \hskip 6pt} \else {\hbox
       {$Q\hskip -5.0pt\vrule height6.0pt depth 0pt\hskip 6pt$}}\fi}
\def\rz{\ifmmode {I\hskip -3pt R} \else {\hbox {$I\hskip -3pt R$}}\fi}
\def\cz{\ifmmode {C\hskip -4.8pt\vrule height5.8pt\hskip 6.3pt} \else
       {\hbox {$C\hskip -4.8pt\vrule height5.8pt\hskip 6.3pt$}}\fi}
\def\au{{\setbox0=\hbox{\lower1.36775ex\hbox{''}\kern-.05em}\dp0=.36775ex\hs
kip0pt\box0}}
\def\ao{{}\kern-.10em\hbox{``}}

\newcommand\Gregbeq{\begin{eqnarray}}
\newcommand\Gregeeq{\end{eqnarray}}

\def\cH{{\cal H}}

\def\h1{{\hat 1}}
\def\h2{{\hat 2}}

\def\3f{\frac{3}{2}}

\newcommand{\roscoff}[1]{}

{\catcode `\@=11 \global\let\AddToReset=\@addtoreset}
\AddToReset{figure}{section}

\DeclareFontFamily{OT1}{rsfs}{}
\DeclareFontShape{OT1}{rsfs}{m}{n}{ <-7> rsfs5 <7-10> rsfs7 <10-> rsfs10}{}
\DeclareMathAlphabet{\mycal}{OT1}{rsfs}{m}{n}

{\catcode `\@=11 \global\let\AddToReset=\@addtoreset}
\AddToReset{equation}{section}
\renewcommand{\theequation}{\thesection.\arabic{equation}}
\newcounter{mnotecount}[section]

\renewcommand{\themnotecount}{\thesection.\arabic{mnotecount}}

\newcommand{\bbR}{\mathbb{R}}

\newcommand{\jlcax}[1]{}
\newcommand{\eean}{\nonumber\end{eqnarray}}

\newcommand{\fourg}{{\mathfrak g }}

\newcommand{\kk}[1]{}

\newcommand{\beq}{\begin{equation}}

\newcommand{\FS}       
                  {F}

\newcommand{\HS} 
       {H_{\mbox{\scriptsize volume}}}

{\ptc{this should be removed in the oberwolfach version}}%

\newcommand{\tD}{{\widetilde D}}
\newcommand{\tL}{{\widetilde L}}

\newcommand{\eeal}[1]{\label{#1}\end{eqnarray}}
\newcommand{\bed}{\begin{deqarr}}
\newcommand{\eed}{\end{deqarr}}
\newcommand{\bedl}[1]{\begin{deqarr}\label{#1}}
\newcommand{\eedl}[2]{\arrlabel{#1}\label{#2}\end{deqarr}}

\newcommand{\tg}{{\widetilde{g}}}

\newcommand{\mcK}{{\mycal K}}

\newcommand{\bel}[1]{\begin{equation}\label{#1}}
\newcommand{\bea}{\begin{eqnarray}}
\newcommand{\bean}{\begin{eqnarray}\nonumber}
\newcommand{\beal}[1]{\begin{eqnarray}\label{#1}}
\newcommand{\eea}{\end{eqnarray}}

\newcommand{\Eq}[1]{Equation~\eq{#1}}

\newcommand{\qed}{\hfill $\Box$}
\newcommand{\qedskip}{\hfill $\Box$\medskip}
\newcommand{\proof}{\noindent {\sc Proof:\ }}
\newcommand{\be}{\begin{equation}}
\newcommand{\eeq}{\end{equation}}
\newcommand{\ee}{\end{equation}}
\newcommand{\beqa}{\begin{eqnarray}}
\newcommand{\eeqa}{\end{eqnarray}}
\newcommand{\beqan}{\begin{eqnarray*}}
\newcommand{\eeqan}{\end{eqnarray*}}
\newcommand{\ba}{\begin{array}}
\newcommand{\ea}{\end{array}}

\newcommand{\hyp}{\mycal S}

\newcommand{\mcM}{{\mycal M}}

\newcommand{\mnote}[1]
{\protect{\stepcounter{mnotecount}}$^{\mbox{\footnotesize
$
\bullet$\themnotecount}}$ \marginpar{
\raggedright\tiny\em
$\!\!\!\!\!\!\,\bullet$\themnotecount: #1} }

\newcommand{\warn}[1]
{\protect{\stepcounter{mnotecount}}$^{\mbox{\footnotesize
$
\bullet$\themnotecount}}$ \marginpar{
\raggedright\tiny\em
$\!\!\!\!\!\!\,\bullet$\themnotecount: {\bf Warning:} #1} }

\newcommand{\Ricc}{\mathrm{Ric}\,}

\newcommand{\N}{\mathbb N}

\newcommand{\eq}[1]{(\ref{#1})}

\newcommand{\ptc}[1]{\mnote{{\bf ptc:}#1}}

\newcommand{\mcC}{{\mycal C}}

\newcommand{\beqar}{\begin{deqarr}}
\newcommand{\eeqar}{\end{deqarr}}

\newcommand{\beaa}{\begin{eqnarray*}}
\newcommand{\eeaa}{\end{eqnarray*}}

\newcommand{\tr}{\mbox{tr}}

\newcommand{\tK}{{\widetilde K}} 

\newcommand{\bethm}{\begin{Theorem}}
\newcommand{\et}{\end{Theorem}}
\newcommand{\bl}{\begin{Lemma}}

\DeclareFontFamily{OT1}{rsfs}{}
\DeclareFontShape{OT1}{rsfs}{m}{n}{ <-7> rsfs5 <7-10> rsfs7 <10-> rsfs10}{}
\DeclareMathAlphabet{\mycal}{OT1}{rsfs}{m}{n}

{\catcode `\@=11 \global\let\AddToReset=\@addtoreset}
\AddToReset{equation}{section}
\renewcommand{\theequation}{\thesection.\arabic{equation}}

{\catcode `\@=11 \global\let\AddToReset=\@addtoreset}
\AddToReset{figure}{section}

{\catcode `\@=11 \global\let\AddToReset=\@addtoreset}
\AddToReset{table}{section}

\renewcommand{\themnotecount}{\thesection.\arabic{mnotecount}}

\begin{document}


\title{Outer trapped surfaces are dense near MOTSs%
\thanks{Preprint UWThPh-2013-17}}
\author{
Piotr T. Chru\'sciel\thanks{Gravitational Physics, University of Vienna, Boltzmanngasse 5, 1090 Vienna, Austria} {}\thanks{Email  {Piotr.Chrusciel@univie.ac.at}, URL {
http://homepage.univie.ac.at/piotr.chrusciel}}
\\ Gregory J. Galloway
\thanks{Department of Mathematics, University of Miami, Coral Gables, FL 33124,  USA}
{}\thanks{Email  galloway@math.miami.edu}  \vspace{0.5em}\\  }
\maketitle

\vspace{-0.2em}

\begin{abstract}
We show that  any
vacuum initial data set  containing a marginally outer trapped surface $S$ and satisfying a ``no KIDs" condition can be perturbed near $S$ so that $S$ becomes strictly outer trapped in the new vacuum initial data set. This, together with the results in~\cite{EGP}, gives a precise sense in which generic initial data containing marginally outer trapped surfaces lead to geodesically incomplete spacetimes.
\end{abstract}

\noindent
\hspace{2.1em} PACS numbers: 04.20.Dw, 02.40.-k

\tableofcontents

\section{Introduction}

In~\cite{EGP} results were obtained concerning the topology of three-dimensional asymptotically flat initial data sets $(\hyp, g, K)$.   It is proven there that if the initial data $3$-manifold $\hyp$ is not diffeomorphic to $\bbR^3$ then it contains an immersed marginally outer trapped surface, cf.~\cite[Theorem 4.1]{EGP}.
Recall that a marginally outer trapped surface (MOTS) in an initial data set $(\hyp, g, K)$ is a closed embedded two-sided hypersurface $S$ in $\hyp$ such that the null expansion $\theta_+$ with respect to one of the two null normal fields $\ell_+$ to $S$  vanishes, $\theta_+ = 0$.  An immersed MOTS is the image under a finite covering map $p: \hat \hyp \to \hyp$ of a MOTS $\hat S$ in the initial data set $(\hat \hyp, \hat g,\hat K)$, where  $\hat g$ and $\hat K$ are the pullbacks via $p$ of $g$ and $K$, respectively.  By  considering trivial covers, we see, in particular, that MOTSs are immersed MOTSs.  Simple examples of immersed MOTSs that are not MOTSs are described in~\cite{EGP}.

A basic point of view espoused in~\cite{EGP} is that conditions on an initial data set that imply the existence of an immersed MOTS (such as Theorem~4.1 in~\cite{EGP}) should be viewed as an `initial data singularity result'. This view is justified by showing that a Penrose type singularity theorem holds for immersed MOTS; cf., Theorem 3.2 (for MOTSs) and Corollary 3.5 (for immersed MOTSs) in~\cite{EGP}.
By covering arguments the proofs can be reduced to the case of a MOTS that separates a Cauchy surface.  In this situation there are several conditions on a MOTS that lead to incompleteness, as indicated in the following proposition (see~\cite{Costa} for related results):

\begin{proposition}\label{sing}
Let $(\mcM,\fourg)$ be a spacetime of dimension $\ge 3$  such that  the following hold.
\begin{enumerate}
\item[(i)] $\mcM$ satisfies the null energy condition, ${\rm Ric}(X,X) = R_{ij}X^iX^j \ge 0$ for all null vectors $X$, and admits a noncompact Cauchy surface $\hyp$.
\item[(ii)]  There exists a closed, connected hypersurface $S \subset \hyp$ that is separating:
$\hyp \setminus S =   U \cup W$ where $U, W \subset \hyp$ are disjoint and open and where
$U$, say, has noncompact closure, such that $S$ is a MOTS with respect to the null normal field $\ell_+$ that points towards $U$.
\item[(iii)]
 \begin{enumerate}
 \item  Either the generic condition holds along \underline{at least one} future inextendible null geodesic $\eta$ emanating from $S$ in the direction of $\ell_+$,
  \item
 or $S$ is a \underline{strictly stable} MOTS,
  \item  or the null second fundamental form $\chi_+$ of $S$ is \underline{not identically zero}.
 \end{enumerate}
\end{enumerate}
Then $(\mcM,\fourg)$ is future null geodesically incomplete.
\end{proposition}

For the sake of completeness, we indicate the proof of Proposition~\ref{sing}, together with relevant definitions, in Section~\ref{s26VIII13.1}.

The generic condition in point $(iii)(a)$ above requires that there be a non-zero tidal acceleration somewhere along a null geodesic $\eta$ as in the statement of the theorem.  More precisely,
it requires that there be a point $p$ on  $\eta$  and a vector $X$ at $p$ orthogonal to $\eta$ such that $g(R(X,\eta')\eta', X)  \ne 0$; cf.~\cite{beem:ehrlich:global,HE}.
The  condition appears in the classical Hawking-Penrose singularity theorem~\cite[Section 8.2, Theorem 2]{HE}, where it is required to hold along all inextendible causal geodesics.  While this condition may seem physically reasonable, mathematically it is rather unsatisfactory, especially from an initial data point of view.  For example, it is not known whether the maximal globally hyperbolic development of {\it generic} vacuum initial data satisfies the generic condition, even in the weak version of point $(iii)(a)$ above.

Similarly, it is not known whether the remaining alternative conditions of point $(iii)$ are satisfied for generic initial data sets with MOTS.

The purpose of this paper is to resolve the issue by showing that,
 given a
vacuum initial data set $(\hyp, g, K)$ that contains a MOTS $S$ and has no local KIDs near $S$, there exists an arbitrarily small perturbation of the data in a neighborhood of $S$ which gives rise to a new vacuum initial data set $(\hyp, g', K')$ in which $S$ becomes strictly outer trapped.

We proceed to a more detailed statement. Given an open set $\Omega$, let $\mcK_\Omega$ denote the kernel of the map
\be
\label{4}
P^*(Y,N)=
\left(
\begin{array}{l}
2(\nabla_{(i}Y_{j)}-\nabla^lY_l g_{ij}-K_{ij}N+\tr K\; N g_{ij})\\
 \\
\nabla^lY_l K_{ij}-2K^l{}_{(i}\nabla_{j)}Y_l+
K^q{}_l\nabla_qY^lg_{ij}-\Delta N g_{ij}+\nabla_i\nabla_j N\\
\; +(\nabla^{p}K_{lp}g_{ij}-\nabla_lK_{ij})Y^l-N \Ricc(g)_{ij}
+2NK^l{}_iK_{jl}-2N (\tr \;K) K_{ij}
\end{array}
\right) \;.
\ee
The   equations $P^*(Y,N)=0$ are called the \emph{vacuum KID equations on $\Omega$}, and their solutions are called Killing Initial Data (KIDs).
We will say that there are \emph{no local KIDs} near a set $S$ if $\mcK_\Omega=\{0\}$ for every open neighborhood $\Omega$ of $S$.
We note that the condition of non-existence of local KIDs is a generic property of several families of vacuum initial data sets, including the asymptotically flat ones~\cite[Theorem~1.3]{ChBeignokids}. In particular, every vacuum asymptotically flat smooth initial data set is the limit of a sequence of smooth such initial data sets, each of which has no local KIDs.%
\footnote{It is not clear that one can always arrange things so that the MOTS persists when perturbing the data. There are, however, situations, where this occurs. The simplest such case is when $S$ is strictly stable. Another one arises when there are topological reasons for the existence of a MOTS.}

We have:

\begin{Theorem}
 \label{T13I13.1}
Let $S$ be a MOTS inside, or at the boundary of, a smooth
initial data set $(\hyp,g,K)$ satisfying the dominant energy condition.
Suppose moreover that there are no local KIDs near $S$, and that $(g,K)$ satisfies either $J^i=0$ near $S$ or $\mu > |J|$ near $S$. Then there exists a family of  initial
data sets  $(\hyp,g_\epsilon,K_\epsilon)$ with $\epsilon >0$  satisfying the dominant energy condition such that
\begin{enumerate}
 \item  $S$ is outer-trapped within  $(\hyp,g_\epsilon,K_\epsilon)$ ,
 \item     $( g_\epsilon,K_\epsilon)$ coincides with $( g ,K )$ outside  an $\epsilon$-neighborhood of $S$, and
 \item   the data $( g_\epsilon,K_\epsilon)$ converge to $( g ,K )$ in the $C^\infty$-topology as $\epsilon$ tends to zero.
\end{enumerate}
If $(g,K)$ is vacuum on a neighbourhood $\Omega$ of $S$ (possibly $\Omega=\hyp$), or if $J^i$ vanishes on $\Omega$, the pairs $( g_\epsilon,K_\epsilon)$  can be chosen to be vacuum on $\Omega$, respectively to satisfy $J^i=0$ on $\Omega$.
\end{Theorem}

The proof of Theorem~\ref{T13I13.1} can be found in Section~\ref{ss31VII13.1}.

It should be pointed out that in Theorem~\ref{T13I13.1} we allow any value of the cosmological constant $\Lambda$.

For simplicity we have assumed smoothness of all fields at hand, but an identical result holds for metrics with finite,
sufficiently high degree of differentiability.

It is clear that an analogue of the vacuum result above holds for specific matter models, e.g.\ Einstein-Maxwell, or Einstein-scalar field equations,
but definite claims seem to require a case-by-case analysis which we have not attempted to carry out.

As should be clear from  the proof of Proposition~\ref{sing},
Theorem~\ref{T13I13.1}  (together with a covering argument from the proof of \cite[Theorem 3.2]{EGP} when the MOTS is not-separating) implies the following:
Let
$(\hyp,g,K)$ be a vacuum initial data set, with $\hyp$ noncompact, and suppose $S$ is a MOTS in $(\hyp,g,K)$, without local KIDs near $S$.  Then there exists an arbitrarily small smooth local perturbation of the initial data to a new vacuum initial data set $(\hyp,g',K' )$ whose maximal globally hyperbolic development is null geodesically incomplete. Or, to put in more colloquial terms, the maximal globally hyperbolic development of the vacuum initial data set $(\hyp,g,K)$ is {\it on the verge} of being null geodesically incomplete, if not already so.

An outline of the proof of Theorem~\ref{T13I13.1} might be in order:
Let $S$ be a  MOTS within an initial data $(\hyp,g,K)$ as in the theorem. We wish to show that there exists an arbitrarily small deformation of the initial data so that $S$ will be outer-trapped.
To accomplish this, as a first step we use the inverse function theorem and the conformal method for solving the constraints  to produce initial data in a small neighborhood of $S$ with unchanged energy-momentum content by imposing suitable  Robin boundary conditions for the conformal factor.
Supposing there are no KIDS on this neighborhood, when $S$ is part of the boundary of $\hyp$,
a gluing smoothes-out the deformation near the other boundary
of the neighborhood, preserving the outer-trapping condition on $S$. When $S$ is an interior submanifold, a subsequent deformation-and-smoothing argument
in the other direction provides the desired initial data.

\section{Definitions}

 Let  $\hyp$ be a spacelike hypersurface in $(\mcM,\fourg)$, and
consider a submanifold  $S\subset \hyp$ of codimension one.
 Assume that $S$
is two-sided in $\hyp$, which means that there exists a globally
defined field $m$ of unit normals to $S$ within $\hyp$. There are
actually two such fields, $m$ and $-m$, we arbitrarily choose one
and call it {\em outer pointing}.
We let $H$
denote the mean extrinsic curvature of $S$ within $\hyp$,
\bel{meancurasdf}
 H:= D_i m^i\;,
\ee
where $D$ is the  Levi-Civita connection
 of the Riemannian metric $g$ induced by $\fourg$ on $\hyp$.

Now, in the discussion above we have assumed that $\hyp$ is embedded into a Lorentzian spacetime $
(\mcM,\fourg)$, so that the objects defined in this section have a clear spacetime meaning. However, no such embedding is actually necessary for our purposes, and a triple $(\hyp,g,K)$ by itself suffices
to define all the quantities relevant for our argument. This will be our framework for Theorem~\ref{T13I13.1}.

We say that $S$ is \emph{outer-future trapped} if
\bel{outertrasdf} \theta_+:=H +(g^{ij}-m^im^j) K_{ij}< 0\;,\ee
where $K$ is the extrinsic curvature tensor of $\hyp$ in $\mcM$. The submanifold $S$ is said to be a \emph{marginally future-outer trapped surface}, or simply \emph{marginally outer trapped surface}, or a MOTS, if the inequality in \eq{outertrasdf} is an equality.
Finally, $S$ is said to be \emph{weakly outer trapped} if strict inequality ``$<$" in \eq{outertrasdf} is replaced by ``$\le$".

We emphasise that marginally trapped surfaces are assumed to be compact  throughout this paper.

\section{Proof of Proposition~\protect\ref{sing}}
 \label{s26VIII13.1}

$(iii)(a)$: We start by noting that the conclusion of Proposition~\protect\ref{sing} follows immediately from  the proof of~\cite[Theorem 3.2]{EGP} when $(iii)(a)$ is replaced by the requirement that the genericity condition holds along \emph{all} null geodesics in the direction of $\ell_+$.
Now, in the absence of the generic condition one has a rigidity statement: If all null geodesics in the direction of $\ell_+$ are future complete then they form a totally geodesic null hypersurface emanating from $S$; see~\cite[Theorem 7.1]{EGP}. The claim easily follows from this.

$(iii)(b)$: Assume that  $S$ is a {\it strictly stable}  MOTS in $\hyp$, by which we mean that the principal eigenvalue of the MOTS stability operator $L_\Sigma$ of \cite[Equation~(1)]{AMS1}
is strictly positive.
 Then $S$ can be perturbed within
$\hyp$ to a strictly outer trapped ($\theta_+ < 0$) surface,
 and  the future null geodesic incompleteness of $(\mcM,\fourg)$ follows (cf.\ \cite{Costa}).

$(iii)(c)$: We define the null second fundamental form
of $S$, say $\chi_+$, as
$$
 \chi_+(X,Y) = g(\nabla_X \ell_+,Y)
 \;,
$$
where  $X,Y$ are tangent to $S$.
Assume that
$\chi_+$  does not  vanish identically.
 Then $S$ can be perturbed to an outer trapped surface in a slight deformation
$\hyp'$ of  $\hyp$:  Indeed, as follows from Raychaudhuri's equation, by pushing $S$ an arbitrarily small amount along the future directed null normal geodesics of $S$ we obtain a surface $S' \subset \hyp'$ which is weakly outer trapped, $\theta_+ \le 0$, with strict inequality at some points. Then $S'$ can be perturbed outward in $\hyp'$ to a strictly outer trapped surface (cf.\ \cite[Lemma 5.2]{AndM2}), and  the result  follows.
\qedskip

We note that when $(iii)(a)$ holds, then  at least one   null geodesic  in the direction of $\ell_+$ is future incomplete. To see that this statement does not hold in general, consider a Cauchy surface $\hyp$ in the extended Schwarzschild spacetime. Then $\hyp$ meets the event horizon with respect to one of the asymptotically flat ends in a MOTS $S$. All outward future directed null normal geodesics to $S$, which correspond to the generators of the horizon, are  future complete.

\section{Conformal deformations}
 \label{s1VIII13.1}

We consider the usual conformal approach to constructing solutions of the constraint
equations. Given initial data  $(\plaing,\plainK)$, let $
\plainL$ denote the trace-free part of $\plainK$, and let $\tau$ be the trace of $\plainK$:
\beq
\plainK^{ij} = \plainL^{ij} + \frac \tau n \plaing^{ij}  \,,
\eeq
where $n = \dim \hyp$.  For any
function $\phi$ and traceless symmetric tensor $\hat L^{ij}$ we set
\beal{31VIII12.1}
 \noplaing_{ij} &= &  \phi^{\frac {4} {(n-2)}}\plaing_{ij}
 \;,
 \\
 \label{conf18aasdf} \noplainL^{ij}&:=&  \phi^{-\frac {2(n+2)} {(n-2)}}(\plainL^{ij}+\hat L^{ij})
 \;,
\\
 \noplainK^{ij} &:=& \noplainL^{ij} + \frac \tau n \noplaing^{ij}
\;,
 \label{conf18basdf}
\eea
and in applications that we have in mind $\hat L^{ij}$ will take the form
\bel{4VIII13.1}
 \hat L^{ij} = D^ i Y^j + D^j Y ^i - \frac 2 n D_k Y^k g^{ij}
 \;.
\ee
We define the momentum vector $\noplainJ$ of $(\noplaing,\noplainK)$ by the formula
\bean
 8 \pi \noplainJ^j & := &
 \noplainD_i(\noplainK^{ij} - \tr_\noplaing \noplainK \noplaing^{ij})
\\
 \nonumber
  &=& \noplainD_i
 \big(\phi^{-\frac {2(n+2)} {(n-2)}}(\plainL^{ij}+\hat L^{ij})\big) - \frac{(n-1)}n \noplainD^j \tau
\\
 & = &
 \phi^{-\frac {2(n+2)} {(n-2)}}  D_i
 (\plainL^{ij}+\hat L^{ij}) -  \frac{(n-1)}n \phi^{-\frac 4 {(n-2)}}  D^j \tau
 \;.
 \label{9I1.4asdf}
\eea

The energy density   $\noplainmu$ of a pair $(\noplaing,\noplainK)$ is defined as
\bea
 16 \pi \noplainmu & : = &
 R(\noplaing) - |\noplainK|_\noplaing^2 + (\tr_\noplaing \noplainK)^2 - 2 \Lambda
 \;,
 \label{9I1.5}
\eea
where $\Lambda$ is the cosmological constant.
If $\noplainmu$ and the matter-current $\noplainJ^i$ have been prescribed (e.g.\  in vacuum $\noplainmu=0=\noplainJ^i$), we obtain a system of equations for $\hat L^{ij}$ and $\phi$:
\bea
 \displaystyle
  D_i\hat L^{ij}
  & = &
 8 \pi \phi^{\frac{2 (n+2)}{(n-2)}}   \noplainJ^j + \frac{(n-1)}n \phi^{\frac{2n}{(n-2)}}  D^j \tau -
   D_i
  \plainL^{ij}
 \;,
 \label{9I1.6errasdf}
\\
 \displaystyle
  \Delta_\plaing \phi - \frac {(n-2)}{4(n-1)}\plainR \phi
  & = &
   -
 \mynottsigmasquare  \phi^{\frac{2-3n}{(n-2)}} + \beta\phi^{\frac{n+2}{(n-2)}}
 -
  \frac{4( n-2) }{ (n-1)}
 \phi^{ \frac{n+2}{(n-2)}} \pi \noplainmu
 \;,
 \phantom{xxx}
 \label{9I1.6asdf}
\eea
where
\bel{conf216asdf}
 \mynottsigmasquare :=\frac
 {(n-2)}{4(n-1)}|\plainL^{ij}+\hat L^{ij}|_\plaing^2\;,\quad  \beta:=\left[\frac{(n-2)}{4n}\tau^2 - \frac
 {(n-2)}{2(n-1)}\Lambda\right]
 \;.
\ee

We will need the   linearisations of the partial differential operators above, including the boundary operators to be described
in Section~\ref{ss11XII12.2}.
Given  $(
\plaing,\plainK)$ with energy and momentum $(\mu,J^i)$ (possibly, but not necessarily, vanishing),
consider a differentiable one-parameter family of solutions $(\phi(\lambda), \hat L^{ij}(\lambda))$  of
\eq{9I1.6errasdf}-\eq{9I1.6asdf} with $\lambda$-independent $(
\noplainmu,\noplainJ^i)=(
\mu,J^i)$, with further  $\phi(0)=1$ and $
\hat L(0)=0$. Set
\bel{11XII12.1}
 \delta \phi := \frac {d \phi(\lambda)}{d \lambda}\Big|_{
\lambda = 0}
\;,
 \qquad
 \delta \plainL^{ij} := \frac {d \hat L^{ij}(\lambda)}{d \lambda}\Big|_{
\lambda = 0}
\;.
\ee
Differentiating  \eq{9I1.6errasdf} and \eq{9I1.6asdf} at $\lambda=0$ we obtain
%
\bea
  D_i
 \delta \plainL^{ij}
 & = &
 \frac{2\delta \phi} {(n-2)} \big(8 \pi (n+1) J^j+(n-1) D^j \tau\big )
 \;,
 \label{9I1.6errasdf2}
\\
  \Delta_\plaing \delta \phi
  & = &
 \left(   \frac {(n-2)}{4(n-1)}\plainR - \frac{2-3n}{(n-2)} \mynottsigmasquare + {\frac{n+2}{(n-2)}}\beta
 - \frac {4\pi (n+2)}{(n-1)} \mu
 \right) \delta \phi
  \nonumber
\\
 && -\frac
 {(n-2)}{2(n-1)} \plainL^{ij} \delta\plainL_{ij}
 \;.
 \label{9I1.6asdf2}
\eea
\Eq{4VIII13.1} leads to fields $\delta  L^ {ij}$ of the form
\bel{3IX12.1ax}
 \delta  L^ {ij} :=  D^ i \delta Y ^ j +  D^j \delta Y ^ i
  - \frac 2 n  D_k \delta Y^k
   \,
    g^{ij}
  \;,
\ee
which turns (\ref{9I1.6errasdf2})-(\ref{9I1.6asdf2}) into an elliptic system for $(\delta \phi, \delta Y)$.

By assumption $\phi=1$ satisfies \eq{9I1.6asdf} with $\noplainmu=\mu$, which gives
\bea
  - \frac {(n-2)}{4(n-1)}\plainR
  & = & -
 \mynottsigmasquare    + \beta - \frac{4\pi (n-2)}{(n-1)}  \mu
 \;.
 \label{9I1.6asdf3}
\eea
It follows that
\eq{9I1.6asdf2} can be rewritten as
\newcommand{\eigenv}{\gamma}
\bea
  \Delta_\plaing \delta \phi -  \underbrace{\left(\frac {4 \left( (n-1) \mynottsigmasquare+ \beta
 \right)}{(n-2)} - \frac {16 \pi}{(n-1)}\mu\right)}_{=:\eigenv}
\delta \phi%
 =
 -\frac
 {(n-2)}{2(n-1)} \plainL^{ij} \delta\plainL_{ij}
 \;.
 \label{9I1.6asdf4}
\eea

Taking into account \eq{3IX12.1ax}, Equations~\eq{9I1.6errasdf2} and \eq{9I1.6asdf4} form a homogeneous linear system of PDEs for $(
\delta  \phi, \delta Y)$.

We will seek solutions $(\phi(\lambda),\hat L^{ij}(\lambda))$  so that the MOTS $S$ becomes future outer trapped for $
\lambda<0$. This will
be done by applying the implicit function theorem to a family of non-linear boundary value problems,
where the solution
is driven by appropriate boundary conditions at $S$, see Section~\ref{ss31VII13.1}. To apply the implicit function theorem it suffices to
show that the map defined by \eq{9I1.6errasdf2} and \eq{9I1.6asdf4} is an isomorphism. Our boundary conditions will be
elliptic, which implies that the problem is Fredholm. They will also be self-adjoint, which reduces the problem to checking that the system \eq{9I1.6errasdf2}, \eq{9I1.6asdf4} has no kernel. The boundary conditions will be introduced in Section~\ref{ss11XII12.2}, while uniqueness
will be established in Section~ \ref{ss11XII12.1}.

\subsection{Conformal deformations of $\theta_+$}
 \label{ss11XII12.2}

Let $\theta_+$ be the future outer-expansion of $S$   in an $n$-dimensional initial data set $(M,\plaing,\plainK)$:
\bel{tMOTSoutertr}
 \theta_+:= H +\underbrace{ K_{ij}( g^{ij}- m^i m^j)}_{=: \theta_{ K}}
 \;.
\ee
Recall that $ K$ is decomposed as $
 K^{ij} =  L^{ij} + \frac {\plaintau } n  g^{ij}$,
where $ L^{ij}$ is traceless, thus $\tr_\plaing \plainK = \plaintau $.

Under the rescaling $ \plaing_{ij} = \phi^{-\frac 4 {(n-2)}} \noplaing_{ij}  $ as in  \eq{31VIII12.1}, we   have
\bea
  \noplainm^i  & = &  \phi^{-\frac {2} {(n-2)}}  m^i
 \;,
\\
 \noplainH
  &  = &  \frac 1 {\sqrt{\det \noplaing_{kl}} } \partial  _i  ( \sqrt{\det \noplaing_{kl}}\,   \noplainm^i)
   =
   \frac {\phi^{-\frac {2 n} {(n-2)}}} {\sqrt{\det \plaing_{kl}} } \partial _ i  ({\phi^{\frac {2n} {(n-2)}}} {\sqrt{\det \plaing_{kl}} }\,  \phi^{-\frac {2} {(n-2)}}  m^i)
   \nonumber
\\
 &   =  &
  \phi^{-\frac {2} {(n-2)}}  \left( H +  \frac{2  (n-1) m (\phi)}{(n-2)\phi}  \right)
 \;,
\\
  \noplaintheta_{  \noplainK}
  & := &
  \noplainK^{ij}(  \noplaing_{ij}-  \noplainm_i  \noplainm_j)
 \nonumber
\\
  & = &
( \noplainL^{ij} + \frac {\plaintau } n  \noplaing^{ij})(  \noplaing_{ij}-  \noplainm_i  \noplainm_j)
 =
 -   \noplainL^{ij}  \noplainm_i  \noplainm_j + \frac {(n-1)} n \plaintau
 \nonumber
\\
  & = &
 - \phi^{-\frac {2n} {(n-2)}}  (L^{ij} + \hat L^{ij})    m_i  m_j  + \frac {(n-1)} n \plaintau
 \;,
\\
  \noplaintheta_+
  & =  &  \noplainH + \noplaintheta_\noplainK
 \nonumber
\\
  & = &
     \phi^{-\frac {2} {(n-2)}}  \left(   \theta_+ +  \frac{ 2 (n-1)   m (\phi)}{(n-2)\phi}  \right)
 + (\phi^{-\frac {2n} {(n-2)}}  - \phi^{-\frac {2} {(n-2)}}) \theta_{ K}
 \nonumber
\\
 &&
  - \phi^{-\frac {2n} {(n-2)}}  \hat L^{ij}  m_i  m_j
    + \frac {(n-1)} n (1-\phi^{-\frac {2n} {(n-2)}})\plaintau
 \;.
\eeal{31VIII12.2}

As already pointed out, we will be using the implicit function theorem to solve our problem, and so we need to
find the linearized boundary operators.
For this, given initial data
$(\plaing, \plainK)$
consider again
a one-parameter differentiable family $(\phi(\lambda), \hat L_{ij}(\lambda)) $
satisfying
$$
 (\phi(0),\hat L_{ij}(0))=(1,0 )
  \;.
$$
As before, let $\delta $ of
a quantity denote  a partial derivative with respect to $\lambda$ at $\lambda=0$.
We find
\bea
  \delta \theta_+
  & = &
     {-\frac {2} {(n-2)}} \,\big (   (     \theta_+  +  {(n-1) }  (  \theta_{ K} -  \plaintau )   ) \delta \phi
   -   {   (n-1)\,   m (\delta \phi)}
   \big)
 \nonumber
\\
 &&
 -   \delta  L^{ij}  m_i  m_j
 \;,
\eeal{31VIII12.3}
and, of course, all the quantities should be evaluated at $S$.
Equivalently,
\bea
    {     m (\delta \phi)}
    + { K}_{ij}  m^i  m^j
      \delta \phi    -\frac {(n-2)} {2 (n-1)}  \delta  L^{ij}  m_i  m_j
   = \frac {(n-2)} {2 (n-1)}\delta \theta_+
\eeal{31VIII12.4}
on $S$.

 \section{Integral identities and uniqueness}
 \label{ss11XII12.1}

We wish to prove uniqueness of the solutions of the linearized boundary problem for the  equations for $(\delta \phi, \delta Y)$ above, with $\delta \theta_+$ prescribed at the boundary,  when the domain of interest is a sufficiently small collar neighborhood of $S$.
This will be done via integration by parts, using standard functional inequalities.
We will be working in an exterior collar neighborhood of $S$, with the normal $m$ of the previous section pointing \emph{towards} $\Omega$ on $S$, and therefore it is convenient to choose $m$ to be the \emph{inwards-pointing} normal to $\partial \Omega$ throughout $\partial \Omega$.

Multiply \eq{9I1.6asdf4} by $\delta \phi$ and integrate by parts over a set with smooth boundary $\Omega$:
\bea  \int _\Omega  | D \delta\phi|^2   + \eigenv
\delta \phi^2
 & = &
  \frac
 {(n-2)}{2(n-1)}\int _\Omega  \plainL^{ij} \delta\plainL_{ij} \delta \phi
 - \int _{\partial \Omega} \delta \phi \,  m (\delta \phi)
 \;.
 \label{2IX12.1}
\eea
(Here, and elsewhere, the Riemannian measure associated to the metric $g$ is used unless explicitly stated otherwise.)
Similarly multiply \eq{9I1.6errasdf2} by $\delta Y_j$ and integrate by parts to obtain
\bea\frac 12
 \int _\Omega
|
 \delta \plainL^{ij}
 |^2
 & = &
 -   \int_ \Omega
 g(\delta Y,Z)  \delta \phi -
 \int _{\partial\Omega}
\delta  Y_j
 \delta \plainL^{ij}  m_i
 \;,
 \label{2IX12.2}
\eea
where $Z^j =  \frac{2} {(n-2)}(8\pi(n+1)J^j + (n-1)D^j \tau)$.

Let $S \subset \partial \Omega $ be a compact hypersurface in a Riemannian manifold $(\hyp,\plaing )$. Let $(x,y^A)$ be a Gauss coordinate system near $S$,
\bel{15VI12.1}
 \plaing  =dx^2 +\plaing _{AB}(x,y^C)dy^A dy^B
 \;,
\ee
with $S$ given as $\{x=0\}$, and $y^A$ being local coordinates on $S$ propagated to a tubular neighborhood of $S^0$ along geodesics normal to $S $.

Let ${\Omega_\epsilon} =  [0,\epsilon]\times S$, for some $\epsilon>0$ that will be chosen shortly.
For reasons that will become apparent in Section~\ref{ss31VII13.1}, we will seek solutions of \eq{9I1.6errasdf2} and \eq{9I1.6asdf4} with the boundary conditions, in the adapted coordinates $(x,y^A)$ near $S=\{x=0\}\subset \partial {\Omega_\epsilon}$,
\bel{3IX12.1}
 \delta Y^A|_S = 0\;,
 \quad
 \partial_x \delta Y^x |_S =0
 \;,
 \quad
   m (\delta \phi)|_{ S} =   \alpha \delta \phi +  {\beta \delta Y^x}+ \eta
  \;,
\ee
where $\alpha$,  {$\beta$}
  and $\eta$ are smooth functions and  where $m=\partial_x$, consistently with our previous notation. On the remaining part of $\partial {\Omega_\epsilon}$ we will assume
\bel{3IX12.1a}
 \delta Y |_{\{\epsilon\}\times S} = 0
 \;,
 \quad
   \delta \phi|_{\{\epsilon\}\times S} = 0
   \;.
\ee
The integral identities \eq{2IX12.1}-\eq{2IX12.2}  and the  boundary conditions \eq {3IX12.1}-\eq{3IX12.1a} with $\eta=0$
 lead to
\bean  \int _{\Omega_\epsilon}  | D \delta\phi|^2   & = &  - \int _{\Omega_\epsilon} \gamma \delta \phi^2
 -
  \int _{S} \alpha
\delta \phi^2
 {-
  \int _{S} \beta
\delta \phi \, \delta Y^x
}
 \nonumber
\\
 &   &
 +
  \frac
 {(n-2)}{2(n-1)}\int _{\Omega_\epsilon}  \plainL^{ij} \delta\plainL_{ij} \delta \phi
 \;,
 \label{2IX12.1x}
\eea
\bea\frac 12
 \int _{\Omega_\epsilon}
|
 \delta \plainL^{ij}
 |^2
 & = &
-   \int_ {\Omega_\epsilon}
 g(\delta Y,Z)  \delta \phi
    + \frac 2 n \int_S H (\delta Y^x)^2
 \;.
 \label{2IX12.2x}
\eea
Adding, we obtain
\bean
 \lefteqn{
   \int _{\Omega_\epsilon}  | D \delta\phi|^2    + \frac 12
|
 \delta \plainL^{ij}
 |^2
 }
 &&
\\
 &=& -  \int _{\Omega_\epsilon}  \gamma \delta \phi^2
 -   \int_ {\Omega_\epsilon}
 g(\delta Y,Z)  \delta \phi+
  \frac
 {(n-2)}{2(n-1)}\int _{\Omega_\epsilon}  \plainL^{ij} \delta\plainL_{ij} \delta \phi
\nonumber
\\
&&
 -
  \int _{S}  \alpha
\delta \phi^2
 {-
  \int _{S} \beta
\delta \phi \, \delta Y^x
}
  + \frac 2 n \int _{S}  H (\delta Y^x)^2
 \;.
  \phantom{xxx}
 \label{2IX12.1xx}
\eea
Let $h$ be the $x$-independent Riemannian metric,
$$
h := dx^2 +\plaing _{AB}(0,y^C)dy^A dy^B  \,.
$$
Recall the Poincar\'e inequality,
\bel{3IX12.3}
 \int_{[0,1]\times S}   \delta \phi^2+ |\delta Y |_h^2
 \le
 C
  \int_{[0,1]\times S}    (\partial_x\delta \phi) ^2 +|\partial_x\delta Y|_h^2
 \;,
\ee
with some constant $C$, for all smooth functions $\delta \phi$   and
vector fields $\delta Y$ satisfying   \eq{3IX12.1a} with $\epsilon=1$ there. To make things clear we wrote $|\delta Y|_h$ and $|\partial_x\delta Y|_h$ for the norm of the vector fields $\delta Y=\delta Y^i\partial_i$ and $\partial_x \delta Y:=(\partial_x \delta Y^k) \partial_k$ in the metric $h$, but in what follows we will simply write $|\delta Y|$ instead of $|\delta Y|_h$,  etc. In \eq{3IX12.3} it is convenient to use the Riemannian measure associated with $h$,
as then by scaling we obtain
\bel{3IX12.4}
 \int_{[0,\epsilon]\times S} |\delta Y|^2 + \delta \phi^2
 \le
 C \epsilon^2
  \int_{[0,\epsilon]\times S} |\partial_x\delta Y|^2 + (\partial_x\delta \phi)^2
 \;,
\ee
with the same constant. However, replacing $C$ by a larger constant if necessary, \eq{3IX12.4}  remains true for all $0<\epsilon\le 1$  when the measure of $h$ is replaced by that of $g$, which we henceforth use until futher notice.

Now, we claim that
\bel{3IX12.5}
 \int_{[0,\epsilon]\times S} |\partial_x\delta Y|^2
 \le
 C
  \int_{[0,\epsilon]\times S} | \delta L|^2
 \;,
\ee
for a constant $C$ which can be chosen independently of $\epsilon$.
To see this, note that
\bel{3IX12.6}
 \int _{\Omega_\epsilon}  \delta L_{ij} \delta   L^{ij} = \int_{\Omega_\epsilon} |  D\delta Y|^2 +  D _i \delta Y_ j   D ^j  \delta Y ^i - \frac 2 n \divvv \delta Y^2
 \;.
\ee
Using our boundary conditions on $\partial {\Omega_\epsilon}$, we find
\beaa
 \int_{\Omega_\epsilon}   D _i \delta Y_ j   D ^j  \delta Y ^i
  & = &
\int_{\partial{\Omega_\epsilon}}  \underbrace{ m_i \delta Y_ j   D ^j  \delta Y ^i}_{=0}
   - \int_{\Omega_\epsilon} \delta Y_ j   \underbrace{
    D _i  D ^j  \delta Y ^i
   }_{=  D ^j  D _i   \delta Y ^i +  R^j {}_k \delta Y^k }
\\
  & = &
\int_{\partial{\Omega_\epsilon}}  \underbrace{ m^j \delta Y_ j   D _i  \delta Y ^i}_{= H(\delta Y^x)^2}
   + \int_{\Omega_\epsilon} ( \plaindiv \delta Y)^2
   - \int_{\Omega_\epsilon}    R_{j  k} \, \delta Y^ j \delta Y^k
 \;.
\eeaa
Inserted into \eq{3IX12.6} this gives
\bel{3IX12.7}
 \int _{\Omega_\epsilon}  \delta L_{ij} \delta   L^{ij} = \int_{\Omega_\epsilon} |  D\delta Y|^2 +  \frac {(n-2)}  n (\plaindivvv \delta Y)^2 - R_{j  k} \, \delta Y^ j \delta Y^k
+  \int_{\partial \Omega_{\epsilon}} H(\delta Y^x)^2
\;
\ee
%


The boundary term above may be handled as follows; here it is convenient to use the product $h$-measure:  With the current boundary conditions we have
\beaa
 \int_{\partial {\Omega_\epsilon}} H (\delta Y^x)^2 & = & \int_{\{0\}\times S} H (\delta Y^x)^2 = -\int_0^\epsilon \frac d {dx} \int_{\{x\}\times S} H (\delta Y^x)^2
\\
  &= &
  \int_{\Omega_\epsilon} \frac {\partial H} {\partial x} (\delta Y^x)^2 + 2 H \delta Y^x \frac{\partial \delta Y^x}{\partial x}
\;.
\eeaa
This can be estimated in modulus by
\bean
{
   C
  \int_{\Omega_\epsilon}  (\delta Y^x)^2 + |\delta Y^x\, \partial_x \delta Y^x|
  }
&
 \le &
  C\big ( \|\delta Y^x \|_{L^2({\Omega_\epsilon})}^2 + \|\delta Y^x \|_{L^2({\Omega_\epsilon})}
    \|\partial_x\delta Y^x \|_{L^2({\Omega_\epsilon})} \big )
\\
    &\le &
     C'(\epsilon^2 + \epsilon)  \|\partial_x\delta Y^x \|_{L^2({\Omega_\epsilon})}^2
\;,
\eeal{29VII13.1}
with some other constant $C'$.  As before, replacing $C'$ by a larger constant if necessary, \eq{29VII13.1} remains true  when the measure of $h$ is replaced by that of $g$.

Using the above estimate and \eq{3IX12.4} in  \eq{3IX12.7}, \eq{3IX12.5} easily follows for
$\epsilon$ small enough. We conclude that
\bel{3IX12.8}
 \int_{[0,\epsilon]\times S} | \delta Y|^2
 \le
 C \epsilon^2
  \int_{[0,\epsilon]\times S} | \delta L|^2
 \;,
\ee
for some constant $C$.



To close the inequalities  it remains to estimate the integrals over $S$ in \eq{2IX12.1xx}. In fact, one of those integrals has just been estimated above, and the  {remaining ones} may be handled in an {essentially} identical manner.
Using \eq{3IX12.4}, \eq{3IX12.8} and these boundary estimates it is straightforward to show now that, choosing $\epsilon$ small enough, the   right-hand side  of \eq{2IX12.1xx} is  dominated by one-half of the left-hand side.
One concludes that $\delta \phi=\delta Y=0$, whence uniqueness.
\qed

\section{Proof of Theorem~\protect\ref{T13I13.1}}
 \label{ss31VII13.1}

We start with a preliminary result:

\begin{Proposition}
  \label{P31VII13.1}
Let $(\hyp,g,K)$  be an
initial data set with energy-momentum $(\mu,J^i)$ and with a marginally  trapped boundary component $S$. Let $\Omega_\epsilon\approx [0,\epsilon]\times S$ denote a  collar neighborhood
 of $S$ extending a $g$-distance $\epsilon$ from $S$. There exist real numbers $\epsilon_0>0$ and $\malph _0(\epsilon)>0$ with the following property:  for all $0<\epsilon<\epsilon_0$ and for all $\malph \in (-\malph _0(\epsilon),\malph _0(\epsilon))$
   there exist smooth fields $(\tg_\malph ,\tK_\malph )$ defined on $\Omega_\epsilon$,   satisfying the constraint equations with $\malph$-independent sources $(\mu,J^i)$, with the future expansion scalar $\tilde \theta_+(\malph) $ of $S$  within $(\tg_\malph ,\tK_\malph )$ satisfying
  \bel{31VII13.21}
  \tilde \theta_+(\malph)=\malph
  \;.
  \ee
Furthermore, for every $k\in \N$ the fields $(\tg_\malph ,\tK_\malph )$ converge to $(g,K)$ in $C^k(\overline \Omega_\epsilon)$ as $\malph $ approaches zero.
\end{Proposition}

\proof
Given a function $\phi$ and a vector field $Y$ let  $(\tg  ,\tK )$ be given by
 \eq{31VIII12.1}-\eq{4VIII13.1}
with, as elsewhere, $
 \plainK^{ij} = \plainL^{ij} + \frac \tau n \plaing^{ij}$.

Let $\beta\in (0,1)$, $k\in \N$. Consider
a Banach manifold $E_1\oplus E_2$ defined as follows: As $E_1$ we take the space of all functions in $C^{k+2,\beta}(\bar \Omega_\epsilon)$
equal to one on $\{\epsilon\}\times S$. As $E_2$ we take the space of
$C^{k+2,\beta}(\bar \Omega_\epsilon)$ sections of the bundle of vectors tangent to $\hyp$ over $\Omega_\epsilon$   satisfying the boundary conditions
\bel{3IX12.1+VII}
  Y^A|_S = 0\;,
 \quad
 \partial_x   Y^x |_S =0
 \;,
 \quad
  Y |_{\{\epsilon\}\times S} = 0
   \;,
\ee
in adapted coordinates as in \eq{15VI12.1}.

Let
$$
 E_1\times E_2 \ni (\phi,Y) \mapsto\mcC(\phi,Y) :=(\tilde \mu, \tilde J^i, \tilde \theta_+)
  \in C^{k,\beta}(\bar \Omega_\epsilon)\times C^{k,\beta}(\bar \Omega_\epsilon) \times
  C^{k+1,\beta}(S)
$$
be the differentiable map which to $(\phi,Y)\in E_1\times E_2 $ assigns the energy and momentum $(\noplainmu,\noplainJ^i)$ of the pair $(\tg,\tK)$, as in \eq{9I1.4asdf}-\eq{9I1.5}, and the outer-future expansion $\tilde \theta_+$ of $S$ in $(\tg,\tK)$.
The derivative of $\mcC$ with respect to $(\phi,Y)$ at $\phi=1$ and $Y=0$ is the operator which has been studied in Section~\ref{ss11XII12.1}, and has been shown to be injective there when $\epsilon$ is chosen small enough. This linear operator is formally self-adjoint. It is Fredholm by standard theory (see, e.g., the a priori estimates of \cite[Theorem~6.30]{GT}, and the comments
 at the end of Section~6.7 there)
hence a linear isomorphism.
We can apply the inverse mapping theorem~\cite[Theorem~5.9]{LangIntroDM} to conclude that $\mcC$ is a diffeomorphism near $(g,K)$.
The pair $(\tg_\malph,\tK_\malph)$ with the required properties is obtained as the  image by $\mcC^{-1}$ of  $(\mu,J^i,s)$.
\qedskip

We are ready to pass to the proof of our main result:

\medskip

{\noindent\sc Proof of Theorem~\ref{T13I13.1}:}
Let $\epsilon>0$.

Suppose, first, that $S$ is a component of the boundary of $\hyp$. Let $\malph_i \nearrow 0$ and, replacing $\epsilon$ by a smaller number if necessary, consider the corresponding sequence of data $(\tg_{\malph_i},\tK_{\malph_i})\to (g,K)$ on a small tubular neighborhood $\Omega_\epsilon$ of $S$, as given by Proposition~\ref{P31VII13.1}. Using the results in~\cite{ChDelay} (compare \cite{ErwannInterpolating}),
the hypothesis that there are no local KIDs near $S$ allows us to glue, for $i$ large enough, $(\tg_{\malph_i},\tK_{\malph_i}) $ with $(g,K)$, in a way such that the resulting initial data set coincides with $(g,K)$ outside of $\Omega_\epsilon$, and with $(\tg_{\malph_i},\tK_{\malph_i}) $  near $S$, and has the same energy-momentum content $(\mu,J^i)$.
In particular we have $\tilde \theta_+=\malph_i<0$, and $S$ is outer-trapped in the new initial data set.
The energy conditions $\mu=0=J^i$, or $\mu \ge 0$ and $J^i=0$, or $\mu > |J|$ will be satisfied (at least for $i$ large enough in the last case) by  construction,
which establishes our claim when $S\subset\partial\hyp$.

Suppose, next, that $S$ is a
submanifold of $\hyp $. The construction just done on the outer side of $S$ makes $S$ outer-trapped in each of the resulting data sets,
 labeled by $\malph_i$. Increasing $i$ if necessary, the construction described in~\cite[Section~8.6]{ChDelay} allows us to find   smooth  pairs $(g_i,K_i)$ which coincide with the data sets already constructed on the outer side of $S$, and which are equal to $(g,K)$ outside of a small inner tubular neighborhood of $S$, with $\mu$ and $J^i$ unchanged.
\qedskip

\noindent{\sc Acknowledgements} PTC acknowledges the friendly hospitality and financial support of University of Miami during part
of work on this paper. His research was further supported in part by Narodowe Centrum Nauki (Poland) under the grant DEC-2011/03/B/ST1/02625,
and by the Austrian Science Fund (FWF) under project   P 23719-N16. GJG acknowledges financial support from the Erwin Schr\"odinger Institute in Vienna, where most of the work on this paper was carried-out.
His research was further supported  by NSF grant DMS-1313724,   and by a grant from the Simons Foundation (Grant No.  63943). 
\def\polhk#1{\setbox0=\hbox{#1}{\ooalign{\hidewidth
  \lower1.5ex\hbox{`}\hidewidth\crcr\unhbox0}}} \def\cprime{$'$}
  \def\cprime{$'$}
\providecommand{\bysame}{\leavevmode\hbox to3em{\hrulefill}\thinspace}
\providecommand{\MR}{\relax\ifhmode\unskip\space\fi MR }
\providecommand{\MRhref}[2]{%
  \href{http://www.ams.org/mathscinet-getitem?mr=#1}{#2}
}
\providecommand{\href}[2]{#2}


\begin{thebibliography}{10}

\bibitem{AMS1}
L.~Andersson, M.~Mars, and W.~Simon, \emph{{Local existence of dynamical and
  trapping horizons}}, Phys.\ Rev.\ Lett. \textbf{95} (2005), 111102,
  arXiv:gr-qc/0506013.

\bibitem{AndM2}
L.~Andersson and J.~Metzger, \emph{{The area of horizons and the trapped
  region}}, Commun.\ Math.\ Phys. \textbf{290} (2009), 941--972,
  arXiv:0708.4252 [gr-qc]. \MR{MR2525646 (2010f:53118)}

\bibitem{beem:ehrlich:global}
John~K. Beem, Paul~E. Ehrlich, and Kevin~L. Easley, \emph{Global {L}orentzian
  geometry}, second ed., Marcel Dekker Inc., New York, 1996.

\bibitem{ChBeignokids}
R.~Beig, P.T. Chru\'{s}ciel, and R.~Schoen, \emph{{KIDs} are non-generic},
  Ann.\ H.\ Poincar\'e \textbf{6} (2005), 155--194, arXiv:gr-qc/0403042.
  \MR{MR2121280 (2005m:83013)}

\bibitem{ChDelay}
P.T. Chru\'{s}ciel and E.~Delay, \emph{On mapping properties of the general
  relativistic constraints operator in weighted function spaces, with
  applications}, M\'em.\ Soc.\ Math.\ de France. \textbf{94} (2003), vi+103,
  arXiv:gr-qc/0301073v2. \MR{MR2031583 (2005f:83008)}

\bibitem{Costa}
I.P. Costa~e Silva, \emph{On the geodesic incompleteness of spacetimes
  containing marginally (outer) trapped surfaces}, Class.\ Quantum Grav.
  \textbf{29} (2012), 235008, 15. \MR{3002870}

\bibitem{ErwannInterpolating}
E.~Delay, \emph{Localized gluing of {Riemannian} metrics in interpolating their
  scalar curvature}, Diff.\ Geom.\ Appl. \textbf{29} (2011), 433--439,
  arXiv:1003.5146 [math.DG]. \MR{2795849 (2012f:53057)}

\bibitem{EGP}
M.~Eichmair, G.J. Galloway, and D.~Pollack, \emph{{Topological censorship from
  the initial data point of view}}, Jour.\ Diff.\ Geom. \textbf{95} (2012),
  389--405, arXiv:1204.0278 [gr-qc]. \MR{3128989}

\bibitem{GT}
D.~Gilbarg and N.S. Trudinger, \emph{Elliptic partial differential equations of
  second order}, Springer, Berlin, 1983.

\bibitem{HE}
S.W. Hawking and G.F.R. Ellis, \emph{The large scale structure of space-time},
  Cambridge University Press, Cambridge, 1973, Cambridge Monographs on
  Mathematical Physics, No. 1. \MR{MR0424186 (54 \#12154)}

\bibitem{LangIntroDM}
S.~Lang, \emph{Introduction to differentiable manifolds}, second ed.,
  Universitext, Springer-Verlag, New York, 2002. \MR{1931083 (2003h:58002)}

\end{thebibliography}
\end{document}